# The Design of a Network-On-Chip Architecture Based On An Avionic Protocol

Ahmed Ben Achballah
INSAT - EPT / LSA, University of Carthage, Tunisia
ahmed.achballah@gmail.com

Slim Ben Saoud
INSAT - EPT / LSA, University of Carthage, Tunisia
slim.bensaoud@gmail.com

*Abstract*—When the Network-On-Chip (NoC) paradigm was introduced, many researchers have proposed many novelistic NoC architectures, tools and design strategies. In this paper we introduce a new approach in the field of designing Network-On-Chip (NoC). Our inspiration came from an avionic protocol which is the AFDX protocol. The proposed NoC architecture is a switch centric architecture, with exclusive shortcuts between hosts and utilizes the flexibility, the reliability and the performances offered by AFDX.

## I. INTRODUCTION

The Network-On-Chip concept is a direct result of the complexity of recent and future System-On-Chips (SoCs). In fact, multiplying the core's number of the same chip has conducted to internal signals communication problems. Conventional buses were not able to manage too many cores with too many signals. Moreover those signals could be heterogeneous in terms of functionality (control, data, and addresses), in terms of speed (different throughputs of internal cores) and we are talking here about multiple clock domains or the most important in terms of priority. Unfortunately, the classic bus architecture like multiple master multiple slave configuration were inefficient to face this multitude of complexity and heterogeneity of such systems.

During the 2000s, the NoC paradigm was introduced by Luca Benini and Giovanni De Micheli [1]. Concerned by the fact that future SoCs with their complexity may not be totally compatible with conventional buses, many researchers have conducted various studies about NoCs [2-12]. The research concerning this area can be classified under 3 principal axes or levels which are the network, the connection and the system levels [13]. By proposing a new architecture we can class our work under the network level [14, 15]. However, when will talk later about strategies, we will explain how this also concerns the connection and system levels.

This paper came after a recent investigation we conducted concerning the use of AFDX protocol as Network-On-Chip [16]. In fact, we have explained our strategy and our inspiration by the AFDX protocol to design a NoC. In this paper, we give an overview of the desired NoC architecture (the switch and the End System) and present the shortcuts idea. At this stage of work, we have only designed and validated the switch core, so we will present only this component in details throughout this manuscript. The next paragraph explains the designed architecture of the entire NoC while paragraph 3 introduces the AFDX protocol. In the fourth paragraph we give the insides of the switch fabric and its internal architecture. Paragraph 5 shows the hardware implementation and validation processes and paragraph 6 discusses the obtained results and exposes the future works.

## II. A BREIF DESCRIPTION OF THE AFDX PROTOCOL

Because of our first inspiration source to design our NoC architecture was the AFDX protocol, the aims of this paragraph is to give an overview about its organization.

In fact AFDX stands for Avionics Full DupleX Switched Ethernet (AFDX) which a standard that describes the physical wiring and the protocol specifications (IEEE 802.3 and ARINC 664 Part 7) for data exchange between avionics subsystems [17]. AFDX is based on Ethernet which is a very mature technology, wildly used and permanently evolving. AFDX offers many enhancements comparing to its antecedent ARINC 429 such as a high-speed data transfer (up to 100 mbps), less wiring and full duplex connection avoiding transmission collisions. We will detail features later in this paragraph.

An AFDX network is a switch centric topology composed by two main elements which are the End System and the switch as shown in Figure 1. The Linking between these two elements is assured by point to point connections.

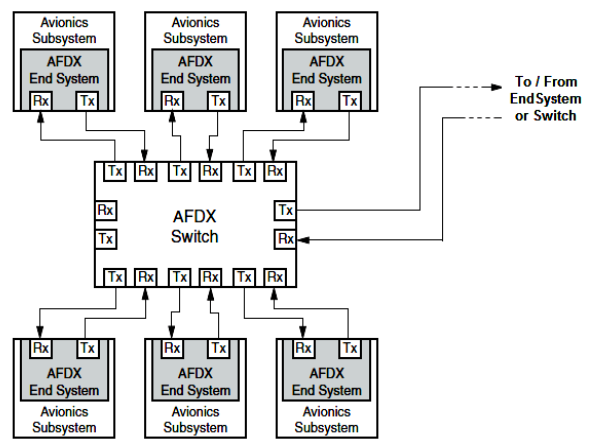

Figure 1. The organization of an AFDX based network

In the following we will give a brief description of the important functionalities of these two components.

## A. The switch

The main function of an AFDX switch is to forward the arriving packets in its Rx ports, originating from End Systems or other switches, to their destination addresses throw its appropriate Tx ports. Internally, this operation is realized by an input/output processing unit. This unit examines every received packet to extract the destination address represented by a virtual link identifier (VLID) and consults a forwarding table to determine the corresponding Tx port to transmit the packet to.

Note that the switch ports (Tx and Rx) are buffered and so are capable of storing multiple incoming/outgoing packets in FIFO (First In, First Out) order. The destination address could be another End System or also another switch depending on the considered architecture.

## B. The End System

In an AFDX network, End Systems play the role of network interfaces to assure communication between avionics subsystems and switches. In fact, they are in charge of receiving messages in there communication ports from avionics equipments, encapsulating them within UDP, IP, and Ethernet headers and placing them on their adequate Virtual Link queue.

The messages packing operation cited before can be realized by one End System for multiple avionics subsystems via multiple ports. However, there are many constraints to respect to allow such configurations. This will be detailed later in virtual link section. After, the queued packets, which are ready to be transmitted, are selected depending on a strategy configured in the virtual link scheduler.

There are many scheduling algorithms to control the selection policy of packets depending on parameters like their length or their importance in all the system. For example, it's obvious that reactor data are more important than displaying a movie to a passenger. Also, this unit may add a sequence number to the frame, per VL basis, starting from 0 to 255 and roll over to 1 (0 is reserved for system reset). This will allow the receiver to check on the received frames from the same VL if they are successive or not. Finally, AFDX frames pass throw a redundancy management unit responsible of replicating and copying them to the physical link.

Finally, as the AFDX protocol was designed for avionic networks, we have to consider many parameters to adapt to our NoC. For example, the redundancy concept is not adopted in our approach. We will expose the differences between the original AFDX protocol and our perception of the designed NoC in a more detailed table.

## III. THE PROPOSED NOC ARCHITECTURE

### A. The Network-On-Chip concept

Network-On-Chip exploits a layered approach to ensure data transfer between IPs, which can be processors, memories, dedicated blocks, etc. Figure shows an OSI 7 layers model and its equivalent in a NoC with its 3 components: Network Interfaces (NI), Physical Links (PL) and Routers (R). The later may be a switch depending on the architecture. In fact, some designer migrate the routing algorithms and strategies back to the network interfaces. In this case, routers are used to dispatch data between IPs and usually called switches. In our proposition, we have used this type of configuration.

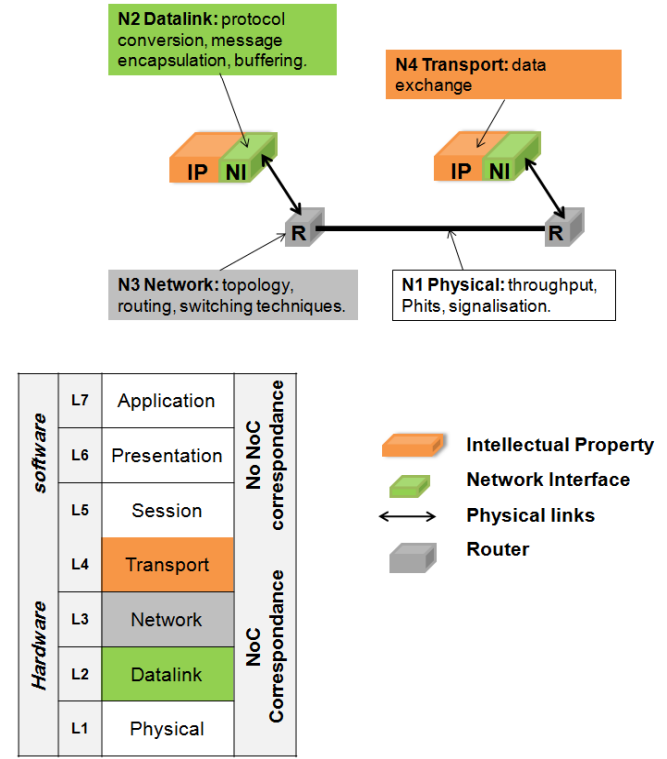

Figure 2. The NoC model and its correspondence with the OSI 7 layers model

### B. The organization of the proposed NoC

However our approach of Network-On-Chip was majorly inspired from AFDX protocol, we have to adopt some intrinsic of that protocol to fill in our project purposes. So, some parameters were utilized, some others were adopted and few others were literally abandoned. Table 1 shows the difference between the original AFDX protocol with its major functionalities and if they were adopted or not in our work.

As we can see in table 1, redundancy concept of AFDX protocol was not adopted. In fact, because AFDX was developed for avionic applications, there are many security standards that in our case are unsuitable or literally impossible to implement. For example, we have adopted parallel connection type instead of the serial one. The main reason is purely technological. In fact, serial connections in SoC are not recommended due to propagation delays, capacitive disturbances and may need repeaters in case of long wires. Another parameter of AFDX that was modified was the address representation.

Again, due to avionic specifications, address representation has to be organized in a special way to fit in desired NoC architecture.

TABLE I. THE DIFFERENCES BETWEEN THE ORIGINAL AFDX PROTOCOL AND THE DESIGNED NOC

| Com. | Functions | | AFDX Protocol | NoC |
|---|---|---|---|---|
| Network | Connection type | | Serial (Macro-network) | Parallel 8 bits (Network On-chip) |
| Network | Redundancy concept | | x | no |
| End System | VL concept | BAG, Lmax | x | x |
| End System | VL concept | Message encapsulation | x | x |
| End System | VL concept | Address representation | x | partial |
| End System | Integrity Checking (SeqN) | | x | x |
| Switch | Filtering | Frame size | x | x |
| Switch | Filtering | FCS field | x | x |
| Switch | Filtering | Destination address | x | x |
| Switch | Policing | BAG | x | No |
| Switch | Policing | Lmax | x | No |
| Switch | Policing | Jitter | x | No |

Although Table 1 shows the differences of all network components, this paper presents only the simulation and implementation results of the switch. This is because this research project is an ongoing work. Results concerning of the End System and all of the NoC will be exposed later in future publications.

## IV. THE SWITCH INTERNAL ARCHITECTURE

The main function of the switch fabric is detailed in figure 3. In fact, the data frame is received on its receiving ports and then processed before to be forwarded or dropped. Each received frame is checked in terms of length and then the FCS is calculated and compared to the frame one. After the verification of the frame length and the FCS field the destination address embedded in the frame is also checked and if found the frame is forwarded in the corresponding port.

The switch forwards frames from its receiving ports to its transmitting ones. Only a filtering operation is processed and there is no modification to any frame. The FCS field is calculated internally and compared to the frame one; there is no recalculation or insertion of a new FCS field inside the switch.

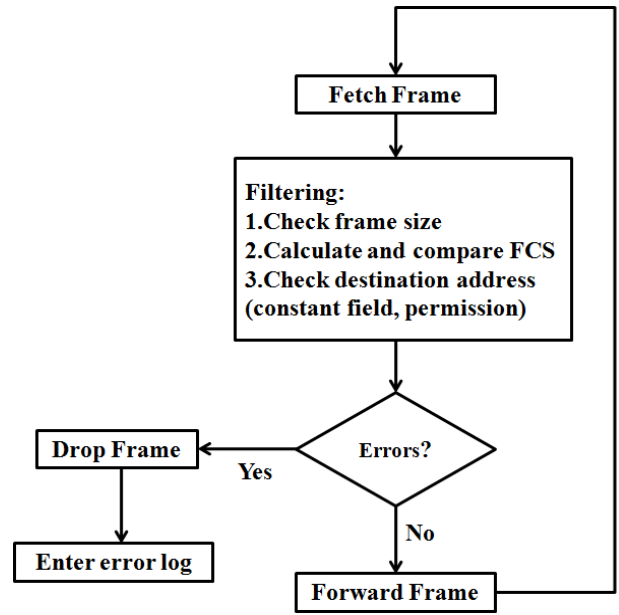

Figure 3. The NoC switch flowchart

The cited functions above are distributed between different cores in our designed hardware module of the switch. The switch module contains 6 cores that execute the switch function that are:

- Test unit
- Switch controller
- CRC module
- Addresses table
- Local storage memory
- Multiplexing matrice

In the following we will explain the role played by each component.

### A. Test unit

This module is implemented on each receiving port of the switch. It is responsible of the detection of new frames. Based on the AFDX protocol, the test unit scans continually the RX ports and compares their state to the preamble seven bytes. When a new frame is detected, a signal is sent to the switch controller.

### B. Local storage memory

When a frame is detected it is globally saved in a local storage memory. It consists on a single port RAM of 2K bytes of size. Knowing that AFDX protocol imposes a max length of a frame of 1518 bytes plus one byte of start frame delimiter and seven bytes preamble, a 1600 bytes sized memory by port basis is sufficient.

### C. CRC module

This core is responsible of calculating the received frame CRC and to compare it to the embedded one in the frame. In both cases, equality or not, the result is signaled to the switch controller.

### D. Addresses table

This component contains the addresses forwarding table of the network. In fact, the data traffic is statically fixed in the network and for each receiving port correspond one or multiple transmitting ports. These modes are called uni- or multi-casting transmission. This configuration is performed by the network designer depending on the application.

### E. CRC controller

The CRC controller manages the workflow of the switch by activating or deactivating one or more switch components. It consists of a finite state machine that organizes the frames forwarding from the receiving ports to the transmitting ones. It is also possible to activate an extra mode which is the broadcasting mode where any received frame if correct is forwarded to all transmitting ports of the switch.

### F. Multiplexing matrice

The main function of this core is similar to classic multiplexer. However, extra features are implemented such as multicasting or broadcasting capabilities. In result, a same input frame can be forwarded to a unique transmitting port, more than one transmitting port or to all ones. This depends on the initial configuration implemented in the addresses table.

## V. HARDWARE IMPLEMENTATION RESULTS

In this paragraph we expose the hardware module of the switch and the obtained results from its simulation and hardware implementation. Figure 4 shows the internal organization of the switch core.

As we can see in figure 4, for each receiving port of the switch a TEST_UNIT is associated. The switch controller receives all indication signals and flags from the rest of the switch components and process the frame in accordance to them.

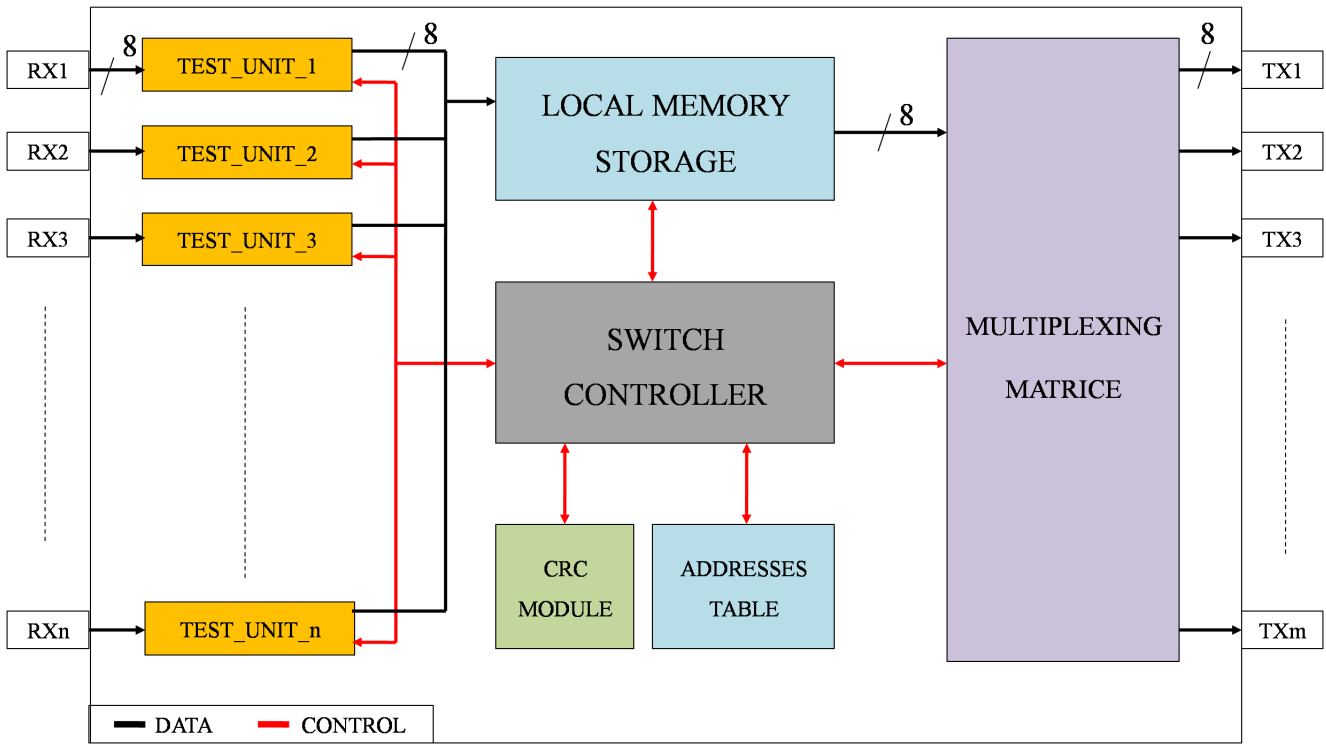

Figure 4. The internal architecture of the NoC switch

The hardware platform utilized to implement and test the switch architecture is the XUP5 development board based on the Xilinx virtex 5 xc5vlx110t FPGA [18]. The hardware design stage using handwritten VHDL and simulation steps were conducted using the ISE design suite 14.4.

Table II shows the hardware synthesis results for the 6 composing elements of the switch. Depending on the configuration, the synthesis result is susceptible to change. If the switch components are studied separately, we can deduce that they are low consuming cores in terms of hardware resources.

As a result, the hardware consumption of all the switch architecture will depend especially on its number of input and output ports. Another advantage of such modular approach is that tests can be executed locally and separately for each component. Starting from already known and fixed set of cores, on-chip verification can be easier.

TABLE II. HARDWARE SYNTHESIS RESULTS OF THE SWITCH CIRCUIT

| Component | Hardware Utilization | | |
|---|---|---|---|
| | Slice Registers | Slice LUTs | Max Frequency (proper to xilinx and depends on technology) |
| SWITCH CONTROLLER | 1130 | 1280 | 170.216 MHz |
| CRC MODULE | 200 | 205 | 280.417 MHz |
| TEST UNIT | 10 | 10 | 452.284 MHz |
| LOCAL STORAGE MEMORY | Use FPGA hardware block RAM | | |
| ADRESSES TABLE | | | |
| MULTIPLEXING MATRICE | | | |

## VI. CONCLUSION AND FUTURE WORK

In this paper a new switch centric architecture of Network-On-Chip is proposed. At this stage of work we have only designed the first component of the NoC which is the switch. The organization of internal switch cores shows a high modularity depending on the desired architecture.

Future works consist of the End System design and the validation of the network with its two components that are the switch and the End System. Moreover, we are planning to add exclusive connections between End Systems for high efficiency communication.

REFERENCES

[1] L. Benini and G. De Micheli, "Networks on Chips: A New SoC Paradigm," *Computer,* pp. 70 - 78, 2002.
[2] E. Bolotin, I. Cidon, R. Ginosar, and A. Kolodny, "QNoC: QoS architecture and design process for network on chip," *Journal of Systems Architecture,* vol. 50, pp. 105–128, 2004.
[3] J. Chan and S. Parameswaran, "NoCEE: Energy Macro-Model Extraction Methodology for Network on Chip Routers," in *IEEE/ACM International Conference on Computer-Aided Design*, 2005, pp. 254 - 259.
[4] J. Chan and S. Parameswaran, "NoCOUT: NoC Topology Generation with Mixed Packet-switched and Point-to-Point Networks," in *Proceedings of the Asia and South Pacific Design Automation Conference*, 2008, pp. 265-270.
[5] V. Soteriou, R. S. Ramanujam, B. Lin, and L.-S. Peh, "A High-Throughput Distributed Shared-Buffer NoC Router," *Computer Architecture Letters* pp. 21 - 24, 2009.
[6] A. Agarwal, C. Iskander, and R. Shankar, "Survey of Network on Chip (NoC) Architectures & Contributions," *Journal of Engineering, Computing and Architecture,* 2009.
[7] P. Zarkesh-Ha, G. B. P. Bezerra, S. Forrest, and M. Moses, "Hybrid network on chip (HNoC): local buses with a global mesh architecture," in *Proceedings of the ACM/IEEE international workshop on System level interconnect prediction*, 2010, pp. 9-14.
[8] Y. He, H. Matsutani, H. Sasaki, and H. Nakamura, "Adaptive Data Compression on 3D Network-on-Chips," *IPSJ Transactions on Advanced Computing Systems,* vol. 5, pp. 13-20, 2012.
[9] M. H. Jabbar, D. Houzet, and O. Hammami, "3D Multiprocessor with 3D NoC Architecture Based on Tezzaron Technology," in *IEEE International 3D System Integration Conference*, 2012.
[10] M. K. P. a. J. C. Hoe, "CONNECT: Re-Examining Conventional Wisdom for Designing NoCs in the Context of FPGAs," in *20th ACM/SIGDA International Symposium on Field-Programmable Gate Arrays* 2012.
[11] M. Coppola, M. D. Grammatikakis, R. Locatelli, G. Maruccia, and L. Pieralisi, *Design of cost-efficient interconnect processing units: Spidergon STNoC*, CRC Press ed., 2008.
[12] *iNoCs*. Available: www.inocs.com
[13] A. Ben Achballah and S. Ben Saoud, "A Survey of Network-On-Chip Tools," *International Journal of Advanced Computer Science and Applications,* vol. 4, pp. 61-67, 2013.
[14] T. Bjerregaard and S. Mahadevan, "A Survey of Research and Practices of Network-on-Chip," *ACM Computing Surveys,* vol. 38, 2006.
[15] E. Salminen, A. Kulmala, and T. D. Hamalainen, "Survey of Network-on-chip Proposals," OCP-IP White paper, 2008.
[16] A. Ben Achballah and S. Ben Saoud, "Investigating The Use of The AFDX Protocol As a Network-On-Chip," in *IEEE International conference on Design & Technology of Integrated Systems in Nanoscale Era*, 2012, pp. 1 - 6.
[17] " Avionics Full Duplex Switched Ethernet (AFDX) Network," Aircraft Data Network Part 7, ARINC 664 Specification.
[18] Xilinx. Available: xilinx.com